\documentclass[prb,preprint,showpacs,preprintnumbers,amsmath,amssymb]{revtex4}

\usepackage[latin1]{inputenc}

\usepackage{graphicx}
\usepackage[tight]{subfigure}
\usepackage{multirow}

\usepackage[thinspace]{SIunits}

\newcommand{\CP}{Co$_{58}$Pt$_{42}$}
\newlength{\LL}
\newlength{\LLL}

\begin{document}

\preprint{APS/123-QED}

\date{\today}

\title{Structural and magnetic properties of CoPt mixed clusters.}

\author{L. Favre}
\altaffiliation[Current address:~]{Laboratoire Matériaux et Microéléctronique de Provence, UMR 6137 CNRS/Universit\'{e} Paul Cézanne Aix-Marseille III, Faculté des Sciences et Techniques de Saint Jérôme, Case 142, 13397 Marseille Cedex 20, France.}
\author{V. Dupuis}
\author{E. Bernstein}
\author{P. M\'{e}linon}
\author{A. P\'{e}rez}
\affiliation{Laboratoire de Physique de la Mati\`{e}re
Condens\'{e}e et Nanostructures, UMR 5586 CNRS/Universit\'{e} Claude Bernard Lyon
1, 69622 Villeurbanne Cedex, France}

\author{S. Stanescu}
\affiliation{ESRF, BP 220, F-38043 Grenoble ,France}

\author{T. Epicier}
\affiliation{Groupe d'\'{E}tude de M\'{e}tallurgie Physique et de
Physique des Mat\'{e}riaux, Institut National des Sciences
Appliqu\'{e}es de Lyon, 69621 Villeurbanne Cedex, France}

\author{J.-P. Simon}
\affiliation{Laboratoire de Thermodynamique et Physicochimie
M\'{e}tallurgiques, UMR 5614 CNRS/INPG/UJF, BP 75, 38402 Saint Martin
D'Heres Cedex, France}

\author{D. Babonneau}
\affiliation{Laboratoire de M\'{e}tallurgie Physique,
 UMR 6630 CNRS/Universit\'{e} de Poitiers, BP 30179, 86962 Futuroscope
Chasseneuil Cedex, France}

\author{J.-M. Tonnerre}
\affiliation{Laboratoire de Cristallographie, CNRS, BP 166, 38042
Grenoble cedex 09, France}

\begin{abstract}
In this present work, we report a structural and magnetic study of mixed \CP\ clusters. MgO, Nb and Si matrix can be used to embed clusters, avoiding any magnetic interactions between particles. Transmission Electron Microscopy (TEM) observations show that \CP\ supported isolated clusters are about $2~\nano\metre$ in diameter and crystallized in the $A1$ fcc chemically disordered phase. Grazing Incidence Small Angle X-ray Scattering (GISAXS) and Grazing Incidence Wide Angle X-ray Scattering (GIWAXS) reveal that buried clusters conserve these properties, interaction with matrix atoms being limited to their first atomic layers. Considering that 60\% of particle atoms are located at surface, this interactions leads to a drastic change in magnetic properties which were investigated with conventional magnetometry and X-Ray Magnetic Circular Dichroïsm (XMCD). Magnetization and blocking temperature are weaker for clusters embedded in Nb than in MgO, and totally vanish in silicon as silicides are formed. Magnetic volume of clusters embedded in MgO is close to the crystallized volume determined by GIWAXS experiments. Cluster can be seen as a pure ferromagnetic CoPt crystallized core surrounded by a cluster-matrix mixed shell. The outer shell plays a predominant role in magnetic properties, especially for clusters embedded in niobium which have a blocking temperature 3 times smaller than clusters embedded in MgO.
\end{abstract}


\maketitle

\section{Introduction}
For several years, the investigation of nanoscale magnetic systems has been stimulating by the demand of high density media storage.
The challenge consists in producing stable ferromagnetic domains in a size range usually facing the superparamagnetic limit~\cite{Neel49,Dorm81,Skum03}. Magnetic clusters, which represent a promising solution, are widely studied~\cite{Sun00,Past89,Past95}. Thermally stable magnetization of clusters requires a high magnetic anisotropy, which can be enhanced by increasing the magneto-crystalline energy (MCE). T$_{x}$Pt$_{y}$ alloys (with T being Co, Fe or Ni) are known to exhibit large MCE due to both hybridization between the transition metal and platinum, and a pronounced crystallographic anisotropy for specific phases~\cite{Xiao04,Chen02,Stap03,Sun04b,Szun99,Tyso96,Uba02}.

In this work, we present a study of crystallographic and magnetic properties of \CP\ magnetic mixed clusters, produced under ultra-high vacuum by laser vaporization and inert gas condensation. Mixed clusters produced by the same technique, have been studied for  Fe-Co, Co-Sm and Co-Ag compounds with larger particle size (\unit{3}{\angstrom} to \unit{12}{\angstrom})\cite{Binn05,Bans05,Dupu04}. The Co-Pt phase diagram~\cite{Mass86,Sanc89}, which is similar to the well known Cu-Au diagram, indicates that four alloy phases exist. At high temperature ($\text{T}\gtrsim 900~\kelvin$) a fcc chemically disordered phase, usually named $A1$, is present for almost any composition. At a lower temperature, Co$_{25}$Pt$_{75}$ and Co$_{75}$Pt$_{25}$ composition lead to a $L1_{2}$ cubic phase, where minority atoms take place in cube corner cell sites, and majority atoms in face centered sites. The Co$_{50}$Pt$_{50}$ composition corresponds to a tetragonal phase, consisting in an alternate stacking of Co and Pt planes. This alloy has been intensively studied because of its high MCE. Recently~\cite{Harp93,Ferr96}, metastable phases have been synthesized using molecular beam epitaxy. They have a short range order and an hexagonal symmetry, leading to a high MCE. As their way of production requires specific thermodynamical conditions different from our set up, they have not been encountered.

In the second section of this article, we will succinctly describe the samples elaboration method and the techniques used to characterize them. The third section is devoted to the structural analysis of both supported and embedded clusters. We focus on the interactions between clusters and matrix, and try to point out the consequences on cluster magnetic properties described in section four.

\section{Experimental setup}

The nanoparticles are prepared using the Low Energy Cluster Beam Deposition (LECBD) technique~\cite{Satt80,Mila90}. Briefly, a Nd: YAG laser is focused on a target-rod with a \mbox{$50$\% Co}, \mbox{$50$\% Pt} atomic composition. The plasma generated at the rod surface is thermalized by a pure He continuous gas flow. The exit nozzle of the nucleation chamber produces an isentropic expansion of the cluster beam. This supersonic beam is collimated by a skimmer before entering the UHV deposition chamber (base pressure: $10^{-9}~\milli\bbar$ reaching $2\cdot 10^{-8} ~\milli\bbar$ during deposition). Because of the isentropic extension, clusters do not fragment upon impact on the substrate~\cite{Pere97}.

Simultaneously, a matrix can be evaporated by electron bombardment under ultra high vacuum (UHV). Both cluster and matrix beams reach the substrate with a $45\degree$ of incidence. To prevent any undesirable clusters interactions, matrix and cluster beam fluxes are tuned to produce very dilute cluster films. Cluster volume concentration is set between 0.1\% and 7\%.

Supported clusters are deposited on carbon coated copper grid and subsequently protected with a thin amorphous silicon layer ($\approx 3~\nano\metre$) to perform ex-situ transmission electron microscopy (TEM) observations. Embedded cluster are co-deposited with a matrix on a (100) silicon wafer, covered by its natural oxidized surface layer.

\medskip
Structure and morphology of isolated \CP\ supported clusters were investigated by High Resolution Transmission Electron Microscopy (HRTEM). Electron diffraction and Rutherford Back Scattering (RBS), with $1.5~\mega\electronvolt$ $\alpha$-particles, were performed on assembled cluster films with a continuous layer equivalent thickness about $2~\nano\metre$, to determine the crystalline phases and mean cluster composition respectively.

Embedded clusters were studied by Small and Wide Angle X-ray Scattering experiments under respectively $0.3\degree$ and $1\degree$ grazing incidence
(GISAXS and GIWAXS) using the 7-circles diffractometer of the D2AM beamline of the European Synchrotron Radiation Facilities (ESRF) laboratory of Grenoble, France. The maximum equivalent cluster thickness is a few nanometers, for a matrix thickness of about few hundred nanometers. The GISAXS intensity has been measured using a Charge Coupled Device (CCD) detector placed at a distance of $670~\milli\metre$ from the sample. The diffraction patterns have been recorded for a photon incident energy set at the absorption Co K edge and below.

Magnetic properties of the clusters were investigated by SQUID using Zero Field Cooled (ZFC) protocols (in an applied field of $10~\milli\tesla$) to determine the average clusters blocking temperature ($T_{\text{B}}$) related to the maximum of the ZFC curve ($T_{\text{m}}$). Average time for magnetic moment measurement is about $10~\second$. As specified in the previous section, clusters are highly diluted in the matrix, avoiding any cluster-cluster magnetic interactions (dipolar, RKKY or superexchange).
Complementary X-ray Magnetic Circular Dichroïsm (XMCD) measurements were performed at the ID8 Beamline (ESRF) using the UHV high-field superconducting magnet setup in a longitudinal geometry and total electron yield (TEY) mode. The absorption spectra were recorded tuning the photon energy across the Co $L_{3,2}$ edges ($778~\electronvolt$ and $793~\electronvolt$, respectively) in an applied magnetic field of $6~\tesla$.

\section{Results}

\subsection{Morphology, structure and composition}
\subsubsection{Supported clusters}

The average \CP\ cluster composition was determined with an incertitude $\Delta=\pm 3\%$. The clusters composition corresponds roughly to the Co$_{50}$Pt$_{50}$ target-rod one. They are lightly Co-enriched, as already observed for this kind of cluster production technique~\cite{Rell04}. The phenomenon may be either due to a predominance of cobalt atoms evaporation upon laser impact or a re-evaporation of weaker linked atoms to evacuate nucleation energy before the isentropic expansion~\cite{Rous95}.

\begin{figure}
\centering
\subfigure[\label{Fig:CoPtSizeDist}]{\includegraphics[width=0.35\linewidth]{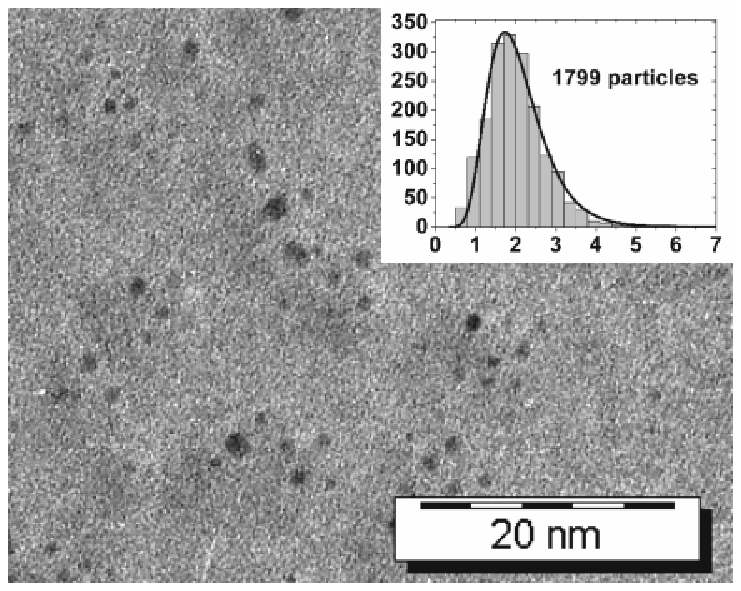}}
\subfigure[\label{Fig:diffCoPt}]{\includegraphics[width=0.26\linewidth]{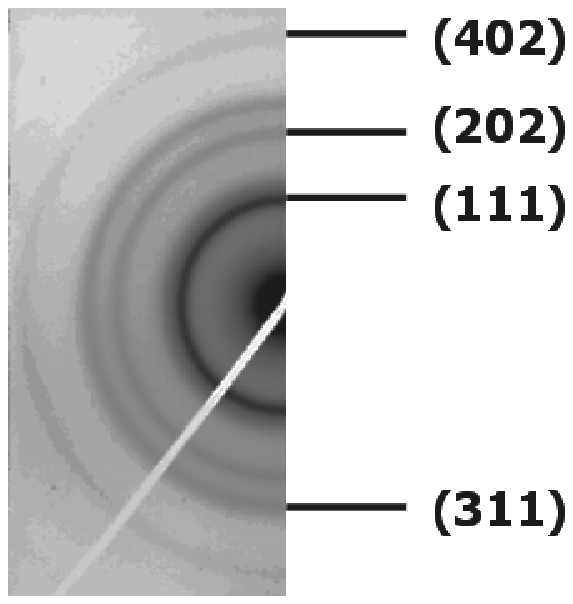}}
\caption{(a)~Micrograph of \CP\ clusters. Inset: \CP\ size distribution (abscissa: Diameter (\nano\metre), ordinate: Number of particles) and log-normal fit curve (mean diameter $D_{m}=2.1~\nano\metre$, dispersion $\sigma =0.35$). (b)~Electron diffraction diagram of \CP\ cluster film. The fcc $A1$~phase diffraction rings are indexed.}
\end{figure}

Figure~\ref{Fig:CoPtSizeDist} displays a TEM picture of \CP\ clusters. The size distribution of the particles shown in the inset is best fitted with a log-normal law, with a mean diameter $D_{m}=2.1~\nano\metre$, and a dispersion $\sigma=0.35$. Inter-reticular distances deduced from electron diffraction (see Fig.~\ref{Fig:diffCoPt}) are reported in table~\ref{Tab:diffCoPt} and compared to $A1$ and $L1_0$ phase inter-reticular distances. The diffraction pattern unambiguously corresponds to (111), (202), (311) and (402) fcc $A1$-phase rings. The (200) ring is too close to the intense (111) ring to be clearly pointed out. The lattice parameter is evaluated to $a=3.80 \pm 0.05~\angstrom$. Therefore a $2.1~\nano\metre$ diameter cluster consists in less than $300$~atoms.

\begin{table}[!hbt]
\centering
\begin{tabular}{|r l|c r|c r|}
 \hline \multicolumn {6}{|c|}{Inter-reticular distances (\angstrom)}\\\hline
 \multicolumn {2}{|c|}{measure}	& \multicolumn {2}{c|}{~~$A1$ phase~~} & \multicolumn {2}{c|}{~~$L1_0$ phase~~}\\\hline
   \multicolumn {2}{|c|}{---}  	&\multicolumn {2}{c|}{---} 	&~~~3.7~~~~& (001) \\
   \multicolumn {2}{|c|}{---}  	&\multicolumn {2}{c|}{---} 	&   2.69   & (110) \\
  ~2.21 & $\pm$\ 0.025 					&~2.21 & (111) 						 	&   2.18   & (111) \\
   \multicolumn {2}{|c|}{---}   & 1.92 & (200) 						 	&   1.91   & (200) \\
   1.36 & $\pm$\ 0.01  					& 1.36 & (220) 							&   1.33   & (202) \\
   1.16 & $\pm$\ 0.01  					& 1.16 & (311) 							&   1.14   & (311) \\
   0.85 & $\pm$\ 0.005 					& 0.86 & (420) 							&   0.95   & (040) \\
   0.77 & $\pm$\ 0.005 					& 0.78 & (422) 							&  \multicolumn {2}{c|}{---} \\ \hline
\end{tabular}
\caption{\label{Tab:diffCoPt}Inter-reticular distances of \CP\ clusters determined from Fig.~\ref{Fig:diffCoPt} and comparison with $A1$ and $L1_{0}$ phase.}
\end{table}

\begin{figure}
\centering
\subfigure[\label{Fig:CoPtHRa}]{\includegraphics[width=0.25\linewidth]{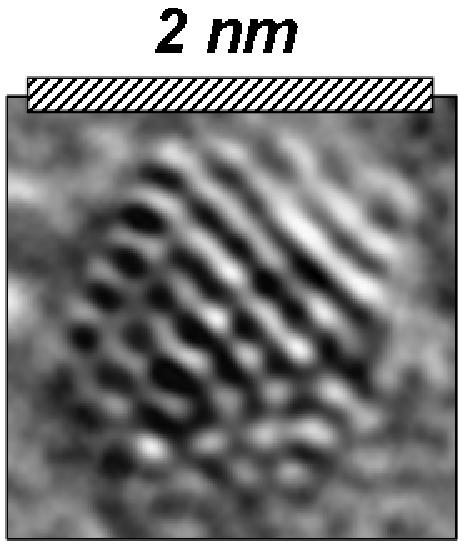}}
\subfigure[\label{Fig:CoPtHRb}]{\includegraphics[width=0.25\linewidth]{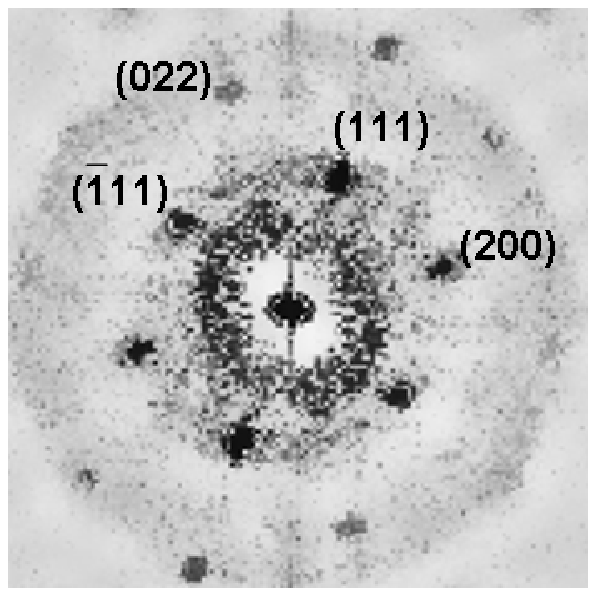}}
\subfigure[\label{Fig:CoPtHRc}]{\includegraphics[width=0.25\linewidth]{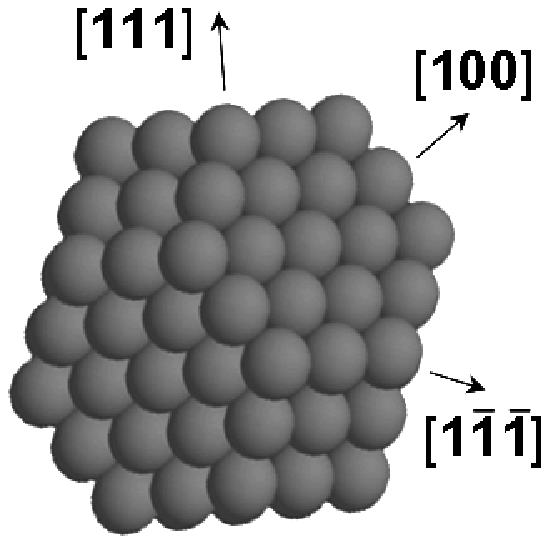}}
\caption{(a)~HRTEM micrograph of \CP\ cluster. (b)~Fourier Transform of the micrograph of the cluster shown in (a). Crystallographic planes are indexed. Inter-reticular distances and angles between planes correspond to $A1$~phase. (c)~Representation of a truncated octahedron containing 201 atoms. The $\left[100\right]$, $\left[111\right]$ and $\left[1\bar{1}\bar{1}\right]$ directions (arrows) are reported.}
\end{figure}

HRTEM observations reveal well faceted and crystallized particles as can be seen on Fig.~\ref{Fig:CoPtHRa}. Inter-reticular distances and angles between atomic planes deduced from Fourier Transform of micrographies, coincide to the $A1$~phase (see Fig.~\ref{Fig:CoPtHRb}). The Fourier Transform of Fig.~\ref{Fig:CoPtHRa} indicates that the particle is oriented along its $\left[01\bar{1}\right]$ direction.

\mbox{Wullf} theorem indicates that the most stable morphology of free fcc clusters in this size range, is either the truncated octahedron or the cuboctahedron, depending on surface tension $\gamma$ of cluster facets~\cite{MacK62,VanH69}.

To our knowledge, surface tension of cobalt-platinum alloys have neither been calculated nor measured. Nevertheless, pure cobalt and pure platinum fcc surface tensions have been calculated (cf. table~\ref{Tab:SurfTens}). For both, truncated octahedron would be the most stable shape as~:
\begin{equation}
\frac{\gamma_{110}}{\gamma_{111}} > \sqrt{\frac{3}{2}} \quad \text{and} \quad \frac{\gamma_{100}}{\gamma_{111}} > \frac{\sqrt{3}}{2}.\label{Eq:Wullf}
\end{equation}
Recently, this result was experimentally confirmed on pure cobalt clusters~\cite{Jame00}.

\settowidth{\LL}{$\approx$ 1.225}
\settowidth{\LLL}{$\approx$ 0.866}
\begin{table}
\centering
\begin{tabular}{|c*{7}{c}|}\hline
Element & $\gamma_{100}$ & $\gamma_{110}$
 & $\gamma_{111}$ &
 $\frac{\gamma_{110}}{\gamma_{111}}$ & $\sqrt{\frac{3}{2}}$
 & $\frac{\gamma_{100}}{\gamma_{111}}$ &$\frac{\sqrt{3}}{2}$\\ \hline
\multicolumn{1}{|c|}{Co fcc} & 2.70 & --- & 2.78 & --- & \multirow{2}{\LL}{$\approx$ 1.225} & 1.03 & \multirow{2}{\LLL}{$\approx$ 0.866} \\ \cline{1-1}
\multicolumn{1}{|c|}{Pt fcc} & 1.378 & 2.009 & 1.004 & 1.231 &  & 1.194 &  \\ \hline
\end{tabular}
\caption {\label{Tab:SurfTens}Surface tension of cobalt
fcc~\cite{Alde94} and Platinum fcc~\cite{Vito98} ($\joule\cdot\rpsquare\metre$). Indexes correspond to family planes.}
\end{table}

Figure~\ref{Fig:CoPtHRc} represents a truncated octahedron oriented along its $\left[01\bar{1}\right]$ direction, containing 201 atoms, with an fcc lattice parameter $a=3.8~\angstrom$. Truncated octahedron is the single morphology coherent with shape and lattice orientation observed on  micrograph~\ref{Fig:CoPtHRa} and other ones. Then, we conclude that \CP\ nanoalloyed clusters adopt truncated octahedron morphology. Considering \mbox{Wullf} theorem, it implies that \CP\ $A1$~phase surface tensions verify relations~\ref{Eq:Wullf}.

\medskip
The clusters produced present a mean diameter close to $2~\nano\metre$. Considering lattice characteristics, such particles have more than 60\% of their atoms located at the surface. Therefore, their properties can be significantly modified when embedded, due to direct interactions with matrix atoms. In the following, we will focus on the structural perturbation induced at the cluster-matrix interface.

\subsubsection{Embedded clusters}

To measure the magnetic properties of non-interacting clusters, we need to work on highly diluted clusters assemblies. They were therefore buried either Nb, MgO or Si matrix, respectively. In order to characterize the evolution of the size, structure and morphology of these embedded clusters, we performed GISAX and GIWAX measurements. High brilliance synchrotron source is mandatory to perform such experiments. Grazing incidence configuration has been used to enhance the sensitivity to clusters signal. The Nb matrix signal being much more intense than clusters one, neither GISAXS nor GIWAXS spectra on samples containing Nb matrix were exploitable.

The anomalous X-rays experiments have been unsuccessful to exacerbate cluster signal: a problem arises from the dimension of the scattering object which is comparable to the phonon wave length of the impinging light ($\approx 2~\nano\metre$). The geometry and the morphology of the particles dominate the effect of absorption at element-specific resonances~\cite{Born99}. We present in the following classical scattering results for a wave length close to the $L_{3}$ edge of cobalt. The GISAXS technique near the angle of total external reflection on very smooth surfaces is well-adapted to analyze the shape, size distribution but also the spatial correlations of buried nano-objects~\cite{Babo05}. We first recorded diffused signal of air, pure Si and pure MgO matrix films without clusters, to isolate cluster spectra during processing.

\begin{figure}[!hbt]
\centering
\includegraphics[width=0.45\linewidth]{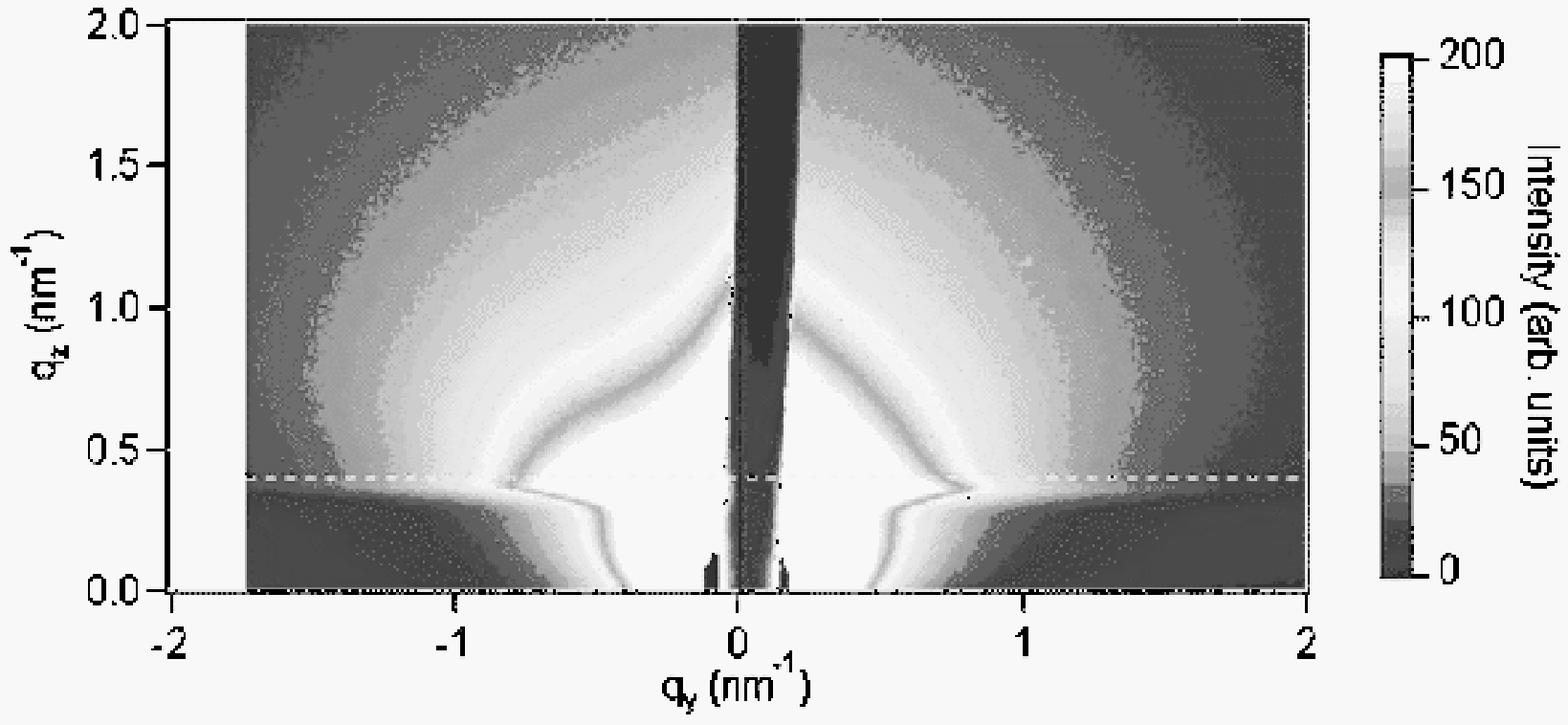}
\caption{\label{Fig:GISAXS_CPMgO_2D}GISAXS signal of the \CP\ clusters in MgO matrix.}
\end{figure}
An asymmetric bulge was observed for Si matrix films, probably related to the $45\degree$ beam incidence during atomic evaporation of the matrix on the substrate. The Kiessig fringes are present in the smooth amorphous Si film $35~\nano\metre$ thick but not in the rough polycrystalline MgO film (see next section for matrices structural properties). For samples containing both clusters and matrix, air and matrix signals were subtracted to the measure. As a typical example, the 2D GISAXS pattern recorded for clusters embedded MgO (see Fig.~\ref{Fig:GISAXS_CPMgO_2D}) exhibits isotropic scattering without interference maximum indicating a random distribution of spherical objects. The corresponding scattered intensity can be described as follows~\cite{Babo05}:
\begin{equation}
I(q_{y},q_{z})=B+k\cdot |T(\alpha _{i})|^2\cdot |T(\alpha _{f})|^2\cdot\!\!\! \int \limits_{0}^{+\infty}\!\!P(q_{y},q_{z},x)f(x,D_{m},\sigma)\ \text{d}x
\end{equation}
where $\vec{q}=(q_y, q_z)$ is the scattering vector corrected for refraction and absorption, $B$ is the background, $k$ a scale factor, $T(\alpha _{i})$ and $T(\alpha _{f})$ are the Fresnel transmission coefficients in incidence and emergence, $P(q_{y},q_{z},x)$ is the form factor of a sphere with diameter $x$ and $f(x,D_{m},\sigma)$ represents the log-normal size distribution function of the clusters depending on the mean diameter $D_{m}$ and the dispersion $\sigma$.

\begin{figure}[!hbt]
 \centering
\subfigure[]{\includegraphics[width=0.45\linewidth]{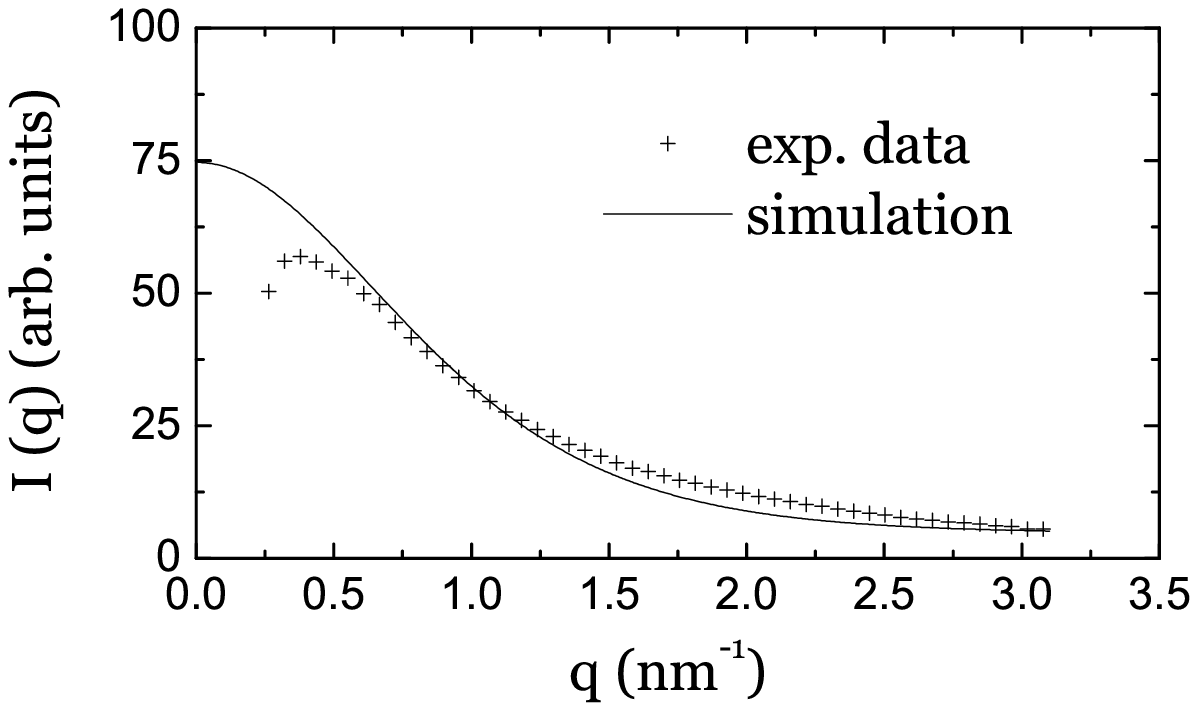}}
\subfigure[]{\includegraphics[width=0.45\linewidth]{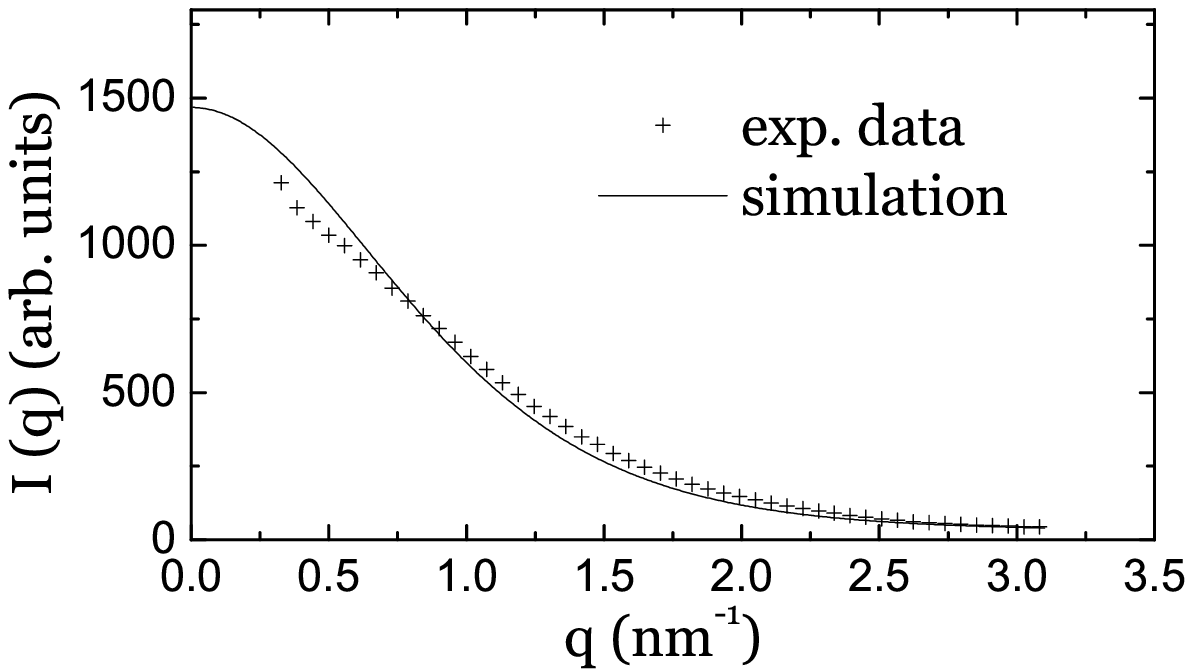}}\\ 
\caption{\label{Fig:GISAXS_CP_matrix}Comparison of GISAXS simulation and measure: \CP\ clusters buried in MgO matrix (a) or Si matrix (b).}
\end{figure}

\begin{figure}[!hbt]
 \centering
\subfigure[]{\includegraphics[width=0.45\linewidth]{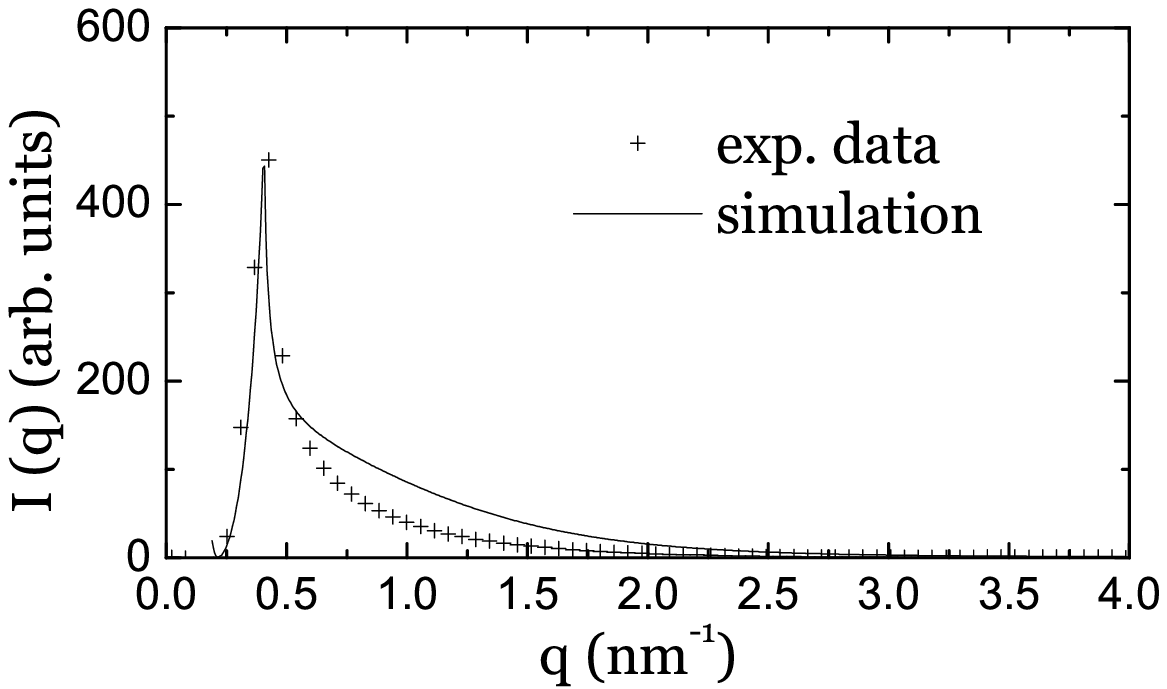}}
\subfigure[]{\includegraphics[width=0.45\linewidth]{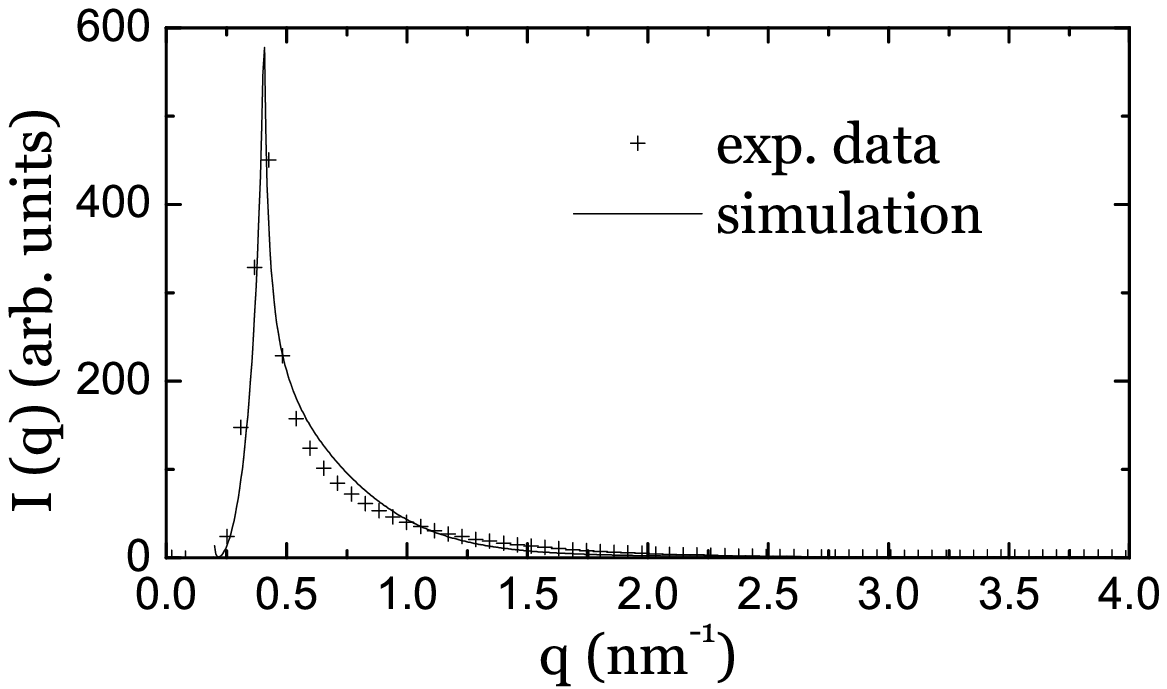}}\\ 
\caption{\label{Fig:GISAXS_CP}Comparison of GISAXS simulation and measure: \CP\ supported clusters with TEM parameters (a) or with a bimodal distribution (b).}
\end{figure}

Fig.~\ref{Fig:GISAXS_CP_matrix} displays measure and simulated curves for $q_{y}=q_{z}$. Simulations performed using the parameters determined from TEM measurements ($D_{m}=2.1~\nano\metre$ and $\sigma=0.35~\nano\metre$) are in agreement with measures for clusters embedded either in MgO or Si matrix. Accordingly, \CP\ clusters shape and size seems not affected by the presence of a matrix. For \CP\ clusters deposited without matrix, the best agreement is obtained for a bimodal repartition $D_{1}=D_{m}$ and $D_{2}=2\cdot D_{m}$ (see Fig.~\ref{Fig:GISAXS_CP}). It is related to coalescence effects: quantity of deposited clusters is higher in GISAXS sample than in TEM sample, to achieve an exploitable signal.

The mean diameter obtained by GISAXS measurements concern the whole particle, whatever its structure is (crystallized or not). As we expect interactions of matrix atoms with cluster ones, we focused on modification of cluster atoms ordering with GIWAXS experiments. We can determine if crystallized volume of the particle is reduced by matrix atoms intrusion and detect appearance of other possible phases. To eliminate Bragg peaks arising from the matrix, we record the diffraction pattern of $35~\nano\metre$-thick Si and MgO films, free of clusters, deposited on a (100) silicon wafer. For both samples, the intense broad peak has been attributed to the native oxide of the substrate, while the amorphous character of the Si film with broad halos and the occurrence of a set of Bragg peaks related to the fcc MgO phase were clearly identified.

\begin{figure}[!hbt]
 \centering
\subfigure[\label{Fig:GIWAXS_CP_MgO}]{\includegraphics[width=0.6\linewidth]{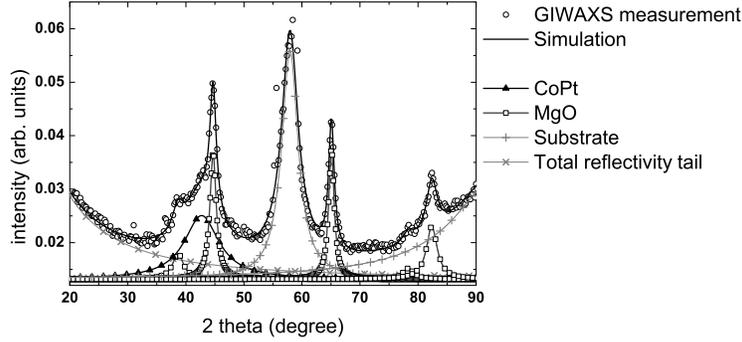}}\\ 
\subfigure[\label{Fig:GIWAXS_CP_Si}]{\includegraphics[width=0.6\linewidth]{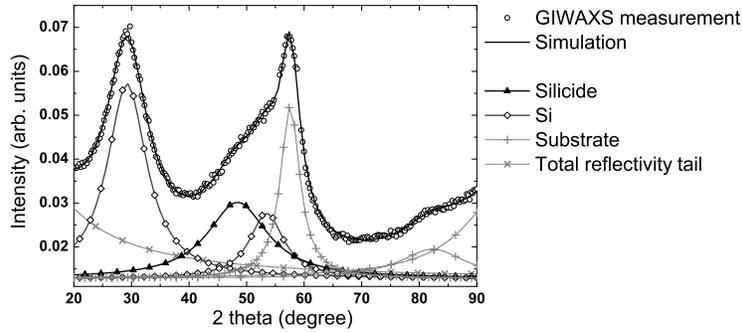}}
\caption{GIWAXS measures and fits for \CP\ clusters embedded in MgO~(a) and Si matrix~(b) respectively.}
\end{figure}

Comparing spectrum of a \CP\ cluster assembled film to free substrate, a Bragg peak arise at $2\theta=42.5\degree$ which corresponds to an inter-reticular distance d$_{111}=2.21~\angstrom$. It is attributed to the (111)-$A1$ CoPt inter-reticular distance. For the sample containing clusters embedded in MgO matrix, this peak is superposed to the fcc MgO diffraction spectrum (see Fig.~\ref{Fig:GIWAXS_CP_MgO} and Tab.\ref{Tab:diffCoPt}). We ca conclude that \CP\ clusters conserve their fcc structure when embedded in MgO. According to Debye-Sherrer formula~\cite{guin64}, we deduce the nanocrystallites mean diameter $\Phi$, taking into account detector resolution ($0\degree 12'$). One finds $\Phi_{CoPt}=1.3~\nano\metre$. This value is smaller than those obtained by TEM and GISAXS experiments ($D_{m}=2.1~\nano\metre$). \CP\ clusters embedded in MgO must be seen as a particle with a mean diameter of $2.1~\nano\metre$ with a $A1$~phase crystallized core with a diameter of $1.3~\nano\metre$. The outer shells are amorphous, due to matrix atoms inclusions.

On the contrary, for \CP\ clusters embedded in a silicon matrix, the characteristic $A1$~phase Bragg peak located at $2\theta=42.5\degree$ vanishes completely. A new broad diffraction peak arises around $2\theta=48\degree$, which corresponds to an inter-reticular distance of about $2~\angstrom$ (see Fig.~\ref{Fig:GIWAXS_CP_Si}), and an average diameter $\Phi_{Co_{x}Si_{y}}=1.4~\nano\metre$. Resolution is not sufficient to precisely identify a phase. The inter-reticular distance can be attributed to various silicide alloys like CoSi, Co$_{2}$Si or Pt$_{2}$Si which are not ferromagnetic. So, the formation of these alloys have a dramatic effect on magnetic properties of \CP\ clusters embedded in Si, as described in the next section.

The structural \CP\ cluster properties are summed up in table~\ref{tab:Sumup}.

\begin{table}[htbp]
\centering
\begin{tabular}{|*{4}{c|}}
\cline{2-4}
 \multicolumn{1}{c|}{}& Supported cluster & Cluster in MgO matrix & Cluster in Si  matrix \\ \hline
Particle Diameter            & $D_{m}=2.1~\nano\metre$ & $D_{m}=2.1~\nano\metre$ & $D_{m}=2.1~\nano\metre$ \\ \hline
Crystallized volume diameter & $D_{m}=2.1~\nano\metre$ & $D_{m}=1.3~\nano\metre$ & $D_{m}=1.4~\nano\metre$ \\ \hline
Phase & CoPt $A1$~phase & CoPt $A1$~phase & silicide phases \\ \hline
\end{tabular}
\caption{\label{tab:Sumup}Structural properties of \CP\ clusters depending on their environment: supported, embedded in MgO or embedded in Si}
\end{table}

\subsection{Magnetic properties}

\begin{figure}[!hbt]
 \centering
\subfigure[]{\includegraphics[width=0.51\linewidth]{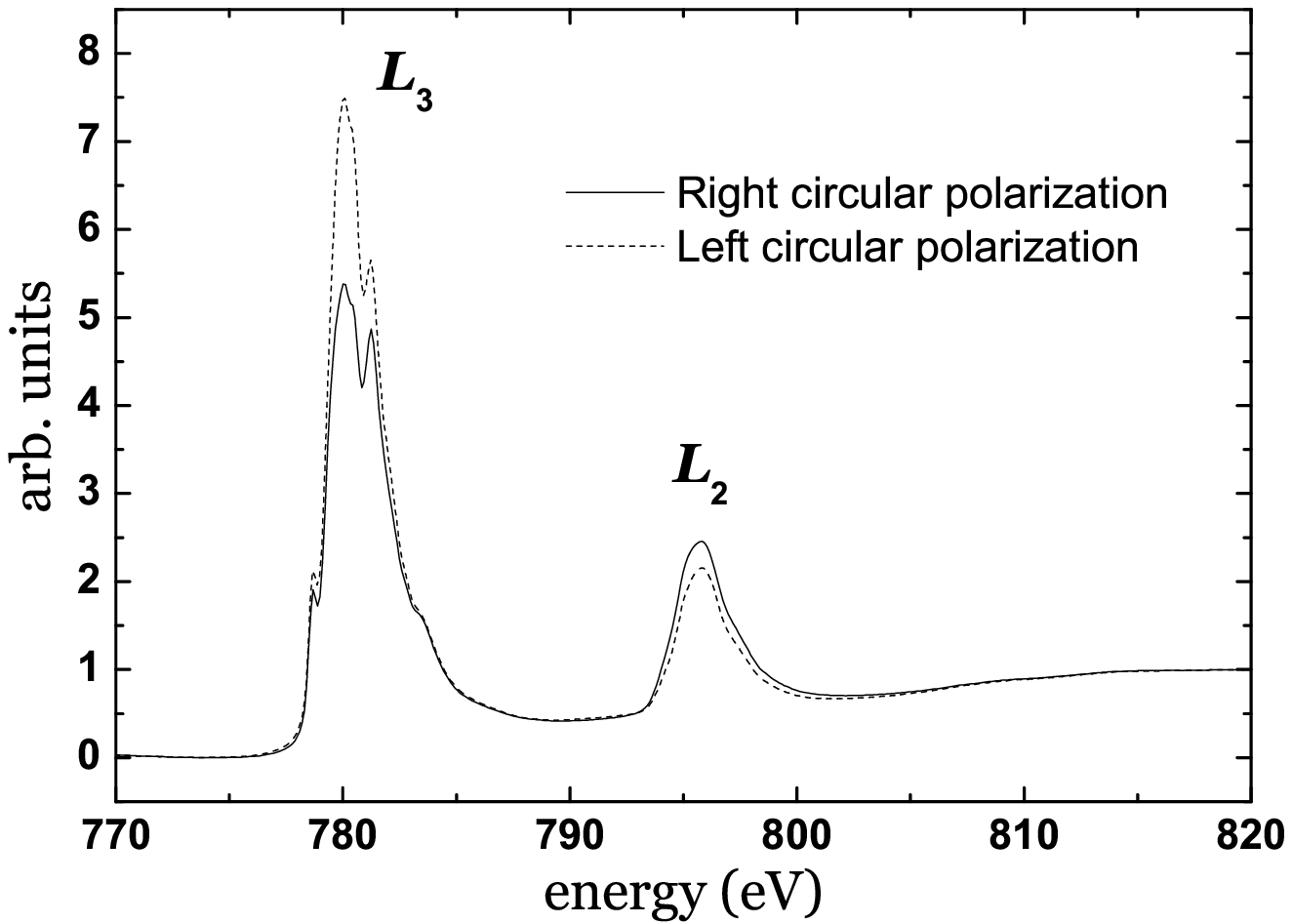}} 
\subfigure[]{\includegraphics[width=0.475\linewidth]{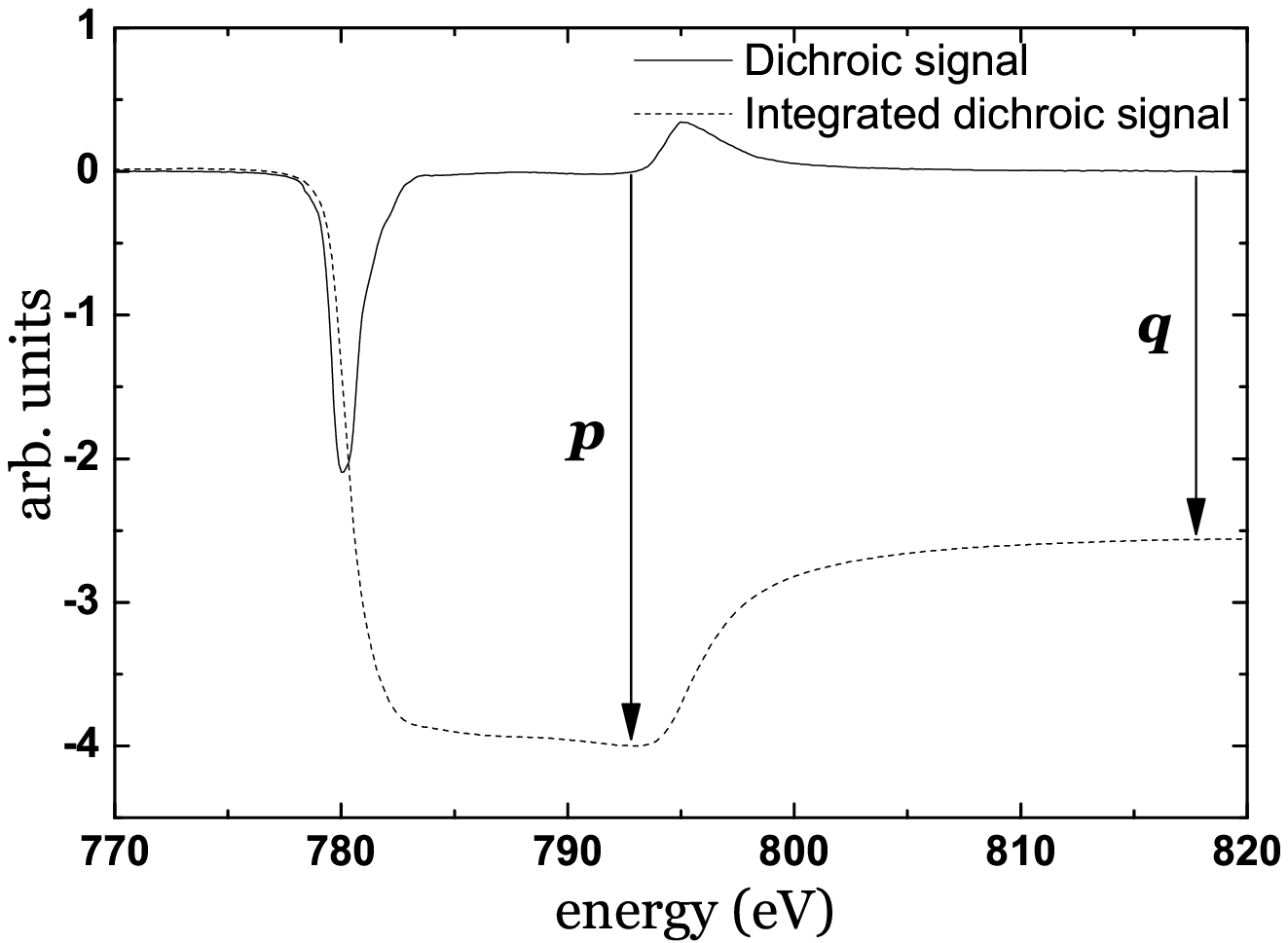}}
\caption{\label{Fig:XMCD_MgO}XMCD measure on CoPt clusters embedded in MgO matrix at the cobalt threshold. (a) Left (plain line) and right (dashed line) polarization spectra. (b) Dichroic signal (plain line) and integrated dichroïc signal (dashed line).}
\end{figure}

To investigate magnetic properties of cobalt atoms, XMCD experiments were performed on \CP\ clusters buried either in MgO, Nb or Si matrix. The dichroïc spectrum is obtained as the difference between two absorption spectra recorded for a parallel and respectively anti parallel configuration between the magnetic field and the polarization (circular left and circular right) of the incoming light. Compared with SQUID measurements, XMCD presents chemical selectivity and enables discriminating between orbital and spin magnetic moments, by using the well known sum rules~\cite{Thol92,Carr93}:
\begin{equation}
m_{orb}=-h \frac{4q}{3r} \mu_{B} \qquad m_{spin}=-h\frac{6p-4q}{r} \mu_{B}
\end{equation}
were $q$, $p$ and $r$ are integrals of the dichroic signal determined experimentally (see Fig.~\ref{Fig:XMCD_MgO}), $h$ the number of holes in the 3d-band, is a quantity usually evaluated for a reference sample or taken from the literature~\cite{Chen95}, and $\mu_{B}$ is the Bohr magneton. These rules have been validated by many experiments~\cite{Wu93,Wu94,Chen95} for transition metals. Here the particles are quasi-spherical and so the dipolar magnetic moment has been considered as null. In this particular system, the evaluation of $h$ is rather difficult: the majority of cobalt atoms are located on the clusters surface, interacting with matrix atoms. The transferability principle using the reference value of $h$ measured on a bulk \CP\ material, may not be relevant. Nevertheless, we will display orbital and spin magnetic moments measured considering $h$ equal to the bulk cobalt value ($h=2.49$): these values must be considered as qualitative information on tendencies. We will also present the orbital-to-spin magnetic moment ratio of the clusters $m_{orb}/m_{spin}$, which do not depend on $h$.

\begin{figure}[!hbt]
 \centering
\subfigure[]{\includegraphics[width=0.49\linewidth]{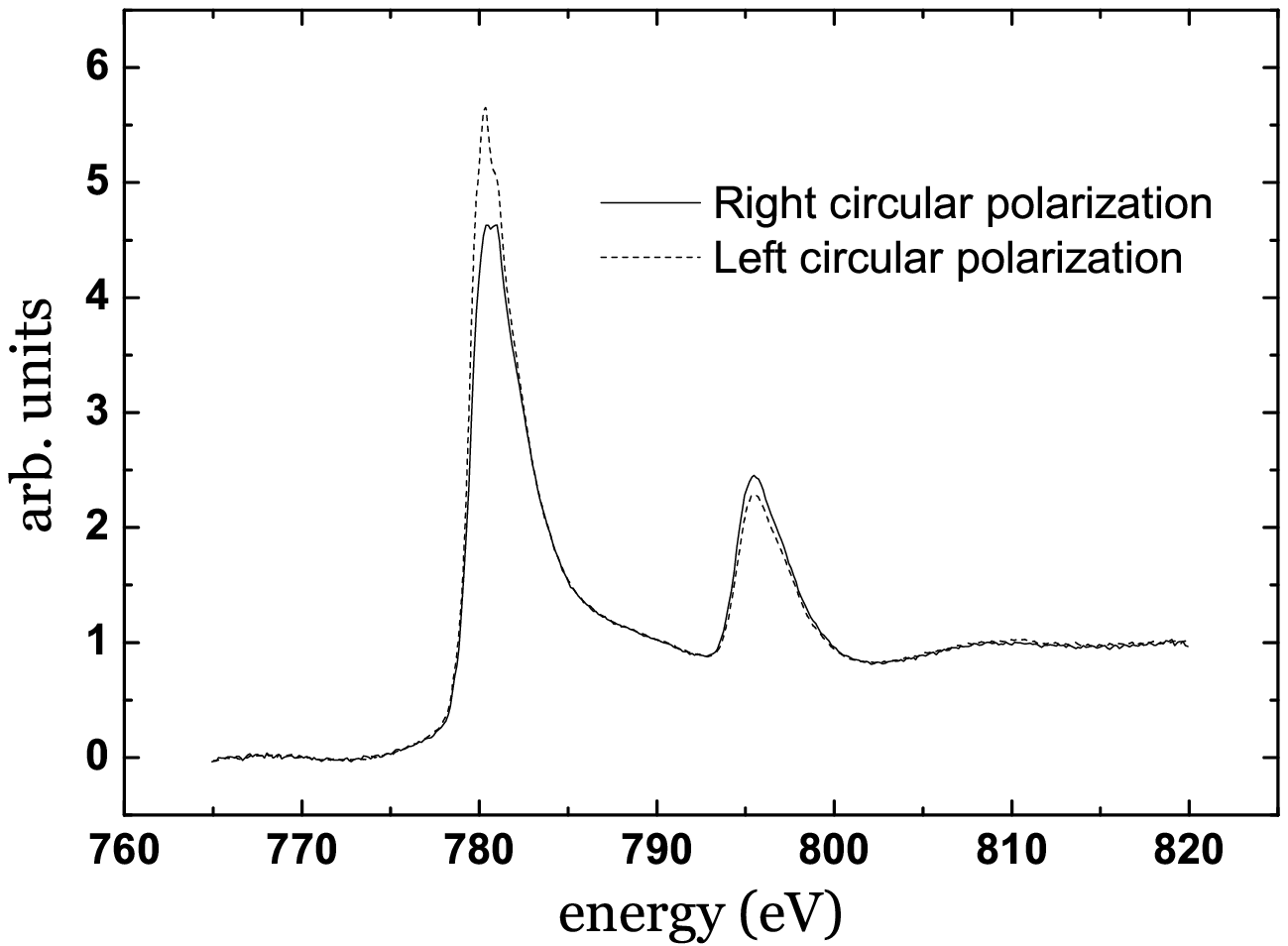}} 
\subfigure[]{\includegraphics[width=0.49\linewidth]{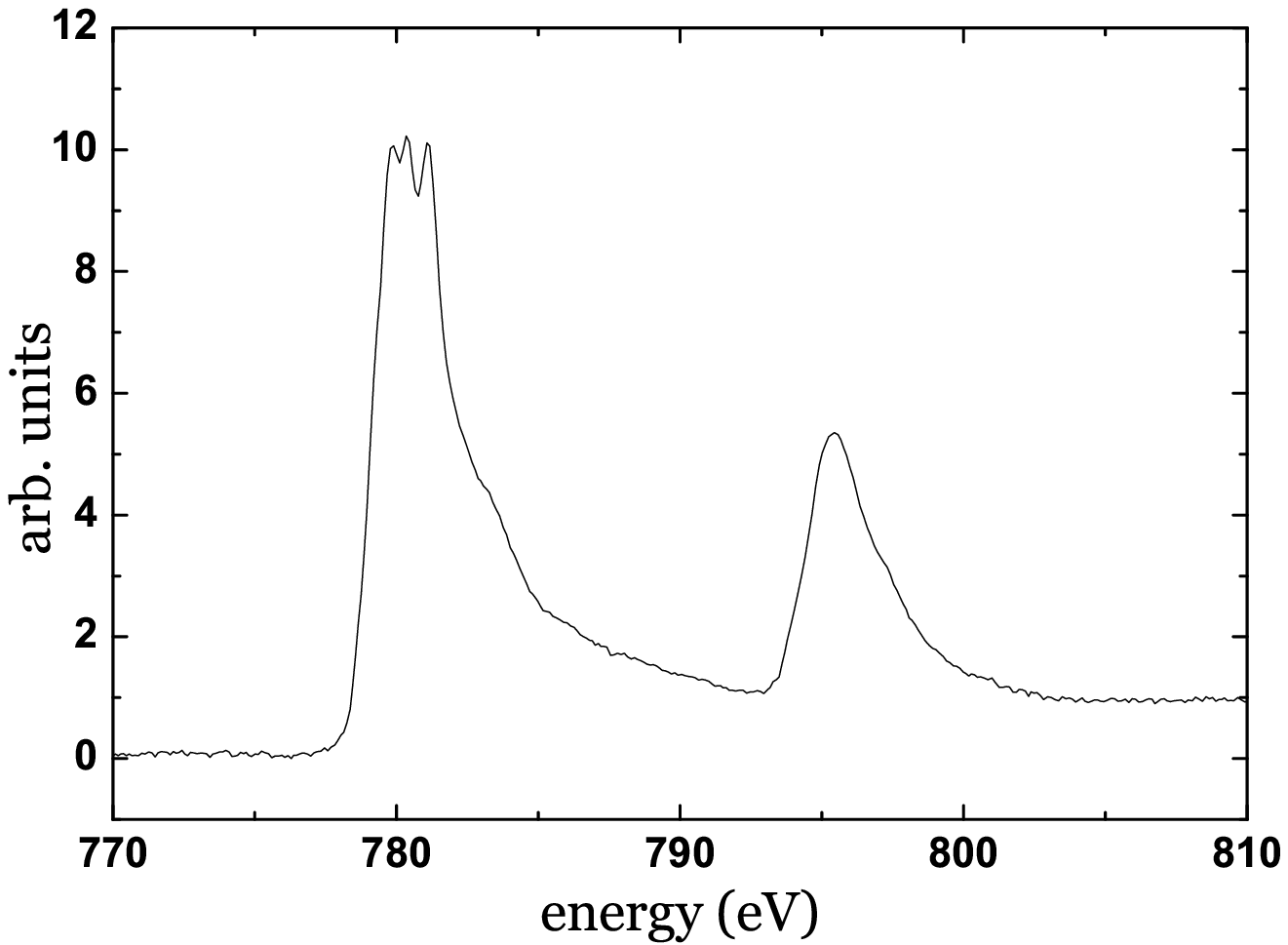}}
\caption{\label{Fig:XMCD_NbSi}XMCD measurements on CoPt clusters embedded in Nb matrix (a) and Si matrix (b) at the cobalt threshold.}
\end{figure}

The absorption signal of the three samples is presented in figures~\ref{Fig:XMCD_MgO} and~\ref{Fig:XMCD_NbSi}. \CP\ clusters embedded in Nb and MgO matrix display a clear dichroic signal. However, the $L_{3}$ threshold absorption signal is different: the usual metallic shape is observed with the Nb matrix, although summit of absorption peak split in several sharp and tiny with the MgO matrix. This multiplet configuration reveals oxidization of some Co atoms that do not, or weakly, contribute to the dichroic signal. 

When \CP\ clusters are embedded in Si matrix, no dichroic signal is observed: the right and left polarization spectra are superposed. One may notice the presence of additional peaks at the $L_{3}$ edge which are attributed to cobalt-silicon bonds in silicides.

\begin{table}[!hbt]
\centering
\begin{tabular}{|c| *{4}{r@{.}l}| *{4}{r@{.}l}|} \cline{2-17}
\multicolumn{1}{c}{} &\multicolumn{8}{|c|}{\CP\ embedded in MgO}& \multicolumn{8}{c|}{\CP\ embedded in Nb}\\ \hline
Temperature & \multicolumn{2}{c}{$m_{\text{orb}}/m_{\text{spin}}$}
& \multicolumn{2}{c}{$m_{\text{orb}}$} & \multicolumn{2}{c}{$m_{\text{spin}}$} & \multicolumn{2}{c|}{$m_{\text{tot}}$} 
& \multicolumn{2}{c}{$m_{\text{orb}}/m_{\text{spin}}$}
& \multicolumn{2}{c}{$m_{\text{orb}}$}
& \multicolumn{2}{c}{$m_{\text{spin}}$}  & \multicolumn{2}{c|}{$m_{\text{tot}}$} \\ \hline
$4~\kelvin$   & ~~~~~~0&25 & 0&15 & 0&60 & 0&74 & ~~~~~~0&31 & 0&06 & 0&20 & 0&26 \\ \hline
$80~\kelvin$  & 0&18 & 0&08 & 0&46 & 0&54 & 0&28 & 0&06 & 0&20 & 0&25 \\ \hline
$300~\kelvin$ & 0&21 & 0&07 & 0&35 & 0&42 & 0&28 & 0&04 & 0&13 & 0&17 \\ \hline
Precision   & $\pm$0&05 & $\pm$0&02 & $\pm$0&05 & $\pm$0&07 & $\pm$0&05 & $\pm$0&02 & $\pm$0&05 & $\pm$0&07 \\ \hline
\end{tabular}
\caption{\label{Tab:XMCD} Orbital and spin magnetic moment per atom ($\mu_{\text{B}}$/at.) deduced from XMCD measurements at the $L_{2,3}$ cobalt thresholds of \CP\ clusters embedded in MgO and Nb matrix respectively. The applied field is 6~\tesla. $m_{\text{orb}}$ and $m_{\text{spin}}$ values are estimated using the theoretical hole number $h=2.49$ in the 3d band of bulk cobalt~\cite{Chen95}. Using the near-saturated magnetization value obtained by SQUID measurement ($215~\kilo\ampere\per\metre$ at $T=300~\kelvin$), one can deduce a platinum magnetic moment per atom $m_{\text{tot}}=0.23\pm0.09~\mu_{\text{B}}$.}
\end{table}

\begin{figure}[!hbt]
\centering
\includegraphics[width=7cm]{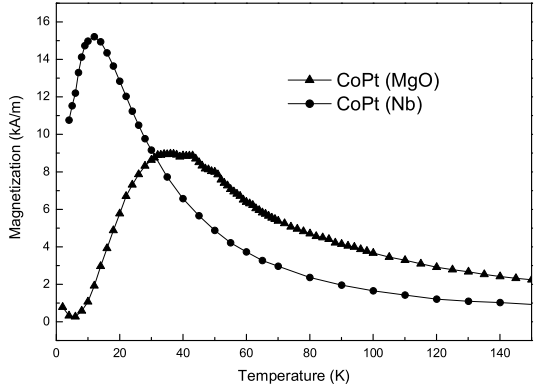}
\caption{\label{Fig:ZFC}Zero Field Cooled measurements of \CP\ clusters embedded in MgO matrix ($T_{\text{m}}\approx 40~\kelvin$) and Nb matrix ($T_{\text{m}}\approx 14~\kelvin$). Lines are guides for the eye.}
\end{figure}

The results are summarized in table~\ref{Tab:XMCD}. The orbital-to-spin magnetic moment ratios are close comparing Nb and MgO embedding matrix. As a comparison, for bulk cobalt $m_{\text{orb}}/m_{\text{spin}}$ is about $0.1$. This enhancement (2 to 3 times higher for clusters) is due to increasing role of symmetry breaking at cluster surface. Moreover matrix atoms can modify the properties of outer shell atoms. A previous study on Co clusters embedded in Nb matrix already demonstrated that a non magnetic Co-Nb alloy layer forms at the cluster surface, reducing the magnetic volume of the nanoparticle~\cite{Jame00}. Assuming the spin moment equals to zero for Co atoms in contact with Nb or O atoms, and equals to the bulk value for other ones, we can deduce the magnetic volume of the cluster ($V_{\text{mag}}$) from the total volume determined by TEM ($V_{\text{TEM}}$):
\begin{equation}
\frac{V_{\text{mag}}}{V_{\text{TEM}}}=\frac{m_{\text{S}}^{\text{exp}}}{m_{\text{S}}^{\text{bulk}}}
\end{equation}
The magnetic to total volume ratio is equal to 0.34 and 0.12 for \CP\ clusters surrounded by MgO and Nb matrix respectively. This result demonstrates that magnetic properties of clusters are strongly dominated by surface characteristics, especially matrix nature. Cluster magnetic volume is reduced by a factor 3 when embedded in MgO, and by a factor 8 when embedded in Nb. This inference is in agreement with the ZFC measurements presented in figure~\ref{Fig:ZFC}. Maximum of the ZFC curve is directly related to the blocking temperature of clusters which depends on their magnetic volume. It is much more reduced for cluster embedded in Nb ($T_{\text{m}}=\unit{14}{\kelvin}$) than in MgO ($T_{\text{m}}=\unit{40}{\kelvin}$).

As demonstrated in the previous section, crystallized diameter (\unit{1.3}{\nano\metre}) of clusters embedded in MgO is smaller than the whole cluster diameter (\unit{2.1}{\nano\metre}). This corresponds to a crystallized to total volume ratio to 0.24, which is close to the magnetic to total volume ratio determined above. Magnetic experiments are in accordance with structural investigations: matrix atoms play a predominant role in cluster properties, as they diffuse in its outer shells. This phenomena is more important in Nb than in MgO matrix. Two mean factors are responsible of the drastic decrease of ferromagnetic properties: the deepness of matrix atom diffusion in the cluster, and the nature interactions between atoms.

\section{Conclusion}

We have presented a study of \CP\ magnetic mixed clusters. They consist in nano-alloyed well faceted particles. Cluster average diameter being close to $2~\nano\metre$, it contains about 300 atoms, with almost 60\% located on the cluster surface. As a consequence, embedding clusters lead to strong interactions with matrix atoms. GISAXS and GIWAXS measurements shows that crystallized volume diameter of \CP\ cluster is reduced from $2.1~\nano\metre$ to $1.3~\nano\metre$ when embedded in MgO. This measurement proves that matrix atoms diffuse in cluster outer shells.

SQUID and XMCD experiments, confirms that cluster magnetic properties strongly depends on the embedding matrix. Blocking temperature and magnetic volume of clusters are smaller in Nb matrix than in MgO matrix. Furthermore, the magnetic volume determined for cluster embedded in MgO is close to the crystallized volume determined by GIWAXS. Embedded cluster can be seen as a pure ferromagnetic CoPt crystallized core surrounded by a cluster-matrix mixed shell.

Two mean factors influence the outer shell properties: its thickness and the nature of cluster-matrix atoms interaction. The first parameter strongly depends on the production conditions and parameters (deposition temperature, matrix evaporation rate, etc.). The second one will meanly influence the magnetic properties. The phase formed can either be diamagnetic, paramagnetic or antiferromagnetic. Further investigations are actually performed on \CP\ clusters embedded in matrix in order to determine the origin of the magnetic anisotropy. This approach should give more information on the role of the cluster-matrix mixed shell. A technical projection would consist in an experimental study on an individual nanoparticle buried in a matrix. Such works are actually in progress in our team, from micro-SQUID and tunneling electronic transport measurements via a single magnetic CoPt particle.

\vspace{3cm}
\begin{acknowledgments}

The authors are indebted to O. Boisron, G. Guiraud and C. Clavier for their continuous technical assistances and developments during LECBD experiments. The authors would like to thank E. Bonet from the Laboratoire Louis N\'{e}el and P. Brooks from the ESRF in Grenoble, France for their collaborations on magnetic measurements. Thanks are also due to CLYME (Consortium Lyonnais de Microscopie \'{E}lectronique) for access to the high-resolution analytical TEM JEOL 2010F. Authors wish to thank S. Rohart (LPMCN, UCBL - CNRS, Villeurbanne, France) and P. Ohresser (UR1, Unité de recherche SOLEIL - CNRS, Gif sur Yvette, France) for the fruitful discussions they had. Finally, the authors gratefully acknowledge support of part of this work from the EU (AMMARE contract no. G5RD-CT 2001-00478 and STREP SFINx no.NMP2-CT-2003-505587).
\end{acknowledgments}

\end{document}